\documentstyle[12pt]{article}       
\begin{document}
\begin{titlepage}
\title{All local quantum states are mixtures of direct products}
\author{Lajos Di\'osi\thanks{E-mail: diosi@rmki.kfki.hu}\\
KFKI Research Institute for Particle and Nuclear Physics\\
H-1525 Budapest 114, POB 49, Hungary\\\\
{\it e-Archive ref.: quant-ph/9506004}}
\date{June 6, 1995}
\maketitle
\begin{abstract}
According to Popescu's recent analysis [Phys. Rev. Lett. {\bf72}, 797
(1994)], {\it nonideal} measurements, rather than ideal ones, may be
emore sensitive to reveal nonlocal correlations between distant parts
of composite quantum systems. The outcome statistics of joint nonideal
measurements on local states should by definition admit local hidden
variable models. We prove that the density operator of a local
composite system must be convex mixture of the subsystems' density
operators. This result depends essentially on a plausible consistency
condition restricting the class of admissible local hidden variable
models.
\end{abstract}
\end{titlepage}

Thirty years ago John Bell's theoretical works \cite{Bel} revealed
peculiar correlations between distant subsystems of composite quantum
systems. Their peculiarity means that they would be impossible between
distant parts of {\it classical} statistical ensembles unless instant
communications are assumed through distances separating them. Since
this latter would contradict to the principle of locality,
the corresponding quantum states are conventionally called
{\it nonlocal} ones. On the contrary: correlations in {\it local}
quantum states do admit classical statistical models called
local-hidden-variable (LHV) models. Hence the notion of locality of
quantum states relies upon what LHV models are.
The issue of whether an arbitrary quantum state admits an LHV model
can not be solved directly. For pure quantum states a trivial
criterion is due to Gisin. All direct product pure states
are local and all entangled pure states are nonlocal \cite{Gis}.
The generalization for mixed quantum states does not exist.
Quite recently, Popescu \cite{Pop} noticed that local tests based
on {\it nonideal} measurements (so--called POVMs \cite{POV}) might
reveal those nonlocalities of certain mixed quantum states which would
remain hidden if only ideal measurements were considered \cite{Wer}.
In the present Letter we follow this idea and prove a simple criterion
of locality valid for pure as well as for mixed states.

As is usual we consider a composite quantum system with Hilbert space
${\cal H}={\cal H}^{(1)}\otimes{\cal H}^{(2)}$ where the two factor
Hilbert spaces belong to the two subsystems. The subsystems are
separated in space. Let density operator $\rho$ stand for the quantum
state of the composite system. Introduce positive-operator-valued
measures (POVMs)
${\cal A}=\{A_\mu\}$ and ${\cal B}=\{B_\nu\}$ on
${\cal H}^{(1)}$ and ${\cal H}^{(2)}$, respectively, where the
observables $A_\mu$ and $B_\nu$ are non-negative Hermitian operators
satisfying the completeness relations
\begin{equation}
\sum_\mu A_\mu=I^{(1)},~~~~~~~\sum_\nu B_\nu=I^{(2)}.
\end{equation}
Here $I^{(1)},I^{(2)}$ stand for the identity operators on
${\cal H}^{(1)}$ and ${\cal H}^{(2)}$, respectively.
For generic POVMs ${\cal A}$ (or ${\cal B}$) the observables
$A_\mu$ (or $B_\nu$) are not necessarily independent of each other
and may satisfy additional linear constraints
\begin{equation}
\sum_\mu f_\mu A_\mu=0,~~~~~~~\sum_\nu g_\nu B_\nu=0
\end{equation}
with real coefficients $f_\mu$ and $g_\nu$.

Let us apply the composite POVM $\{A_\mu\otimes B_\nu\}$ to the state
$\rho$ of the composite system. The expectation value of the local
observable $A_\mu\otimes B_\nu$ takes the standard form \cite{POV}
\begin{equation}
\langle A_\mu\otimes B_\nu \rangle = tr (\rho A_\mu\otimes B_\nu).
\end{equation}
We say that the state $\rho$ admits local hidden variable (LHV) model
with nonideal measurements if, for each pair $\mu,\nu$, the above
expectation value can be expressed as a weighted sum of products of
conditional expectation values of $A_\mu$ and $B_\nu$, respectively:
\begin{equation}
\langle A_\mu\otimes B_\nu \rangle=\sum_\lambda p_\lambda
                                E^{(1)}_{\cal A}(A_\mu\vert\lambda)
                                E^{(2)}_{\cal B}(B_\nu\vert\lambda).
\end{equation}
The $\lambda$'s parametrize the just mentioned "conditions" and have
given the name {\it local hidden variables}\/;
$p_\lambda$ is their normalized probability distribution:
$p_\lambda\geq0$ and $\sum_\lambda p_\lambda=1$. The expectation
values should satisfy the conditions
\begin{equation}
0\leq E^{(1)}_{\cal A}(A_\mu\vert\lambda)\leq\Vert A_\mu\Vert,~~~~~
0\leq E^{(2)}_{\cal B}(B_\nu\vert\lambda)\leq\Vert B_\nu\Vert,
\end{equation}
and
$\sum_\mu E^{(1)}_{\cal A}(A_\mu\vert\lambda)
=\sum_\nu E^{(2)}_{\cal B}(B_\nu\vert\lambda)=1$.
Observe that, by assumption, the
expectation value of $A_\mu$ should not depend on the POVM ${\cal B}$
chosen at distance and, vice versa, the choice of the POVM ${\cal A}$
will not influence the expectation values of the $B_\nu$'s.

We have hitherto constructed a natural extension of the ordinary
LHV models based on ideal (von Neumann) measurements, by allowing
generalized (i.e. nonideal) measurements through POVMs. Now we
introduce the notion of {\it consistent}\/ LHV models: we require the
conditional expectation values on the RHS of Eq.~(4) to be consistent
with the operator constraints (2):
\begin{equation}
\sum_\mu f_\mu E^{(1)}_{\cal A}(A_\mu\vert\lambda)=0,~~~~~~~
\sum_\nu g_\nu E^{(2)}_{\cal B}(A_\nu\vert\lambda)=0
\end{equation}
for all $\lambda$, whenever the Eqs.(2) hold.

It is {\it plausible}\/ to call a given quantum state $\rho$ of the
composite system {\it local}\/ if and only if {\it consistent}\/ LHV
models are admitted  for each choice of the POVMs ${\cal A}$ and
${\cal B}$. In the remaining part of our Letter we prove a simple
mathematical consequence of such a narrower definition of locality.

For simplicity's sake, we assume that
$dim{\cal H}^{(1)}=dim{\cal H}^{(2)}=2$.
First we construct a subtle POVM ${\cal A}$ on ${\cal H}^{(1)}$.
We introduce the pure state projectors
\hbox{$A_{\bf m}={1\over2}(1+\bf{m}\mbox{\boldmath$\sigma$})$} where
${\bf m}$ is unit vector in $R^3$ and \mbox{\boldmath$\sigma$} is the
3-vector of Pauli-matrices. Let us make the following specific choice
for the POVM on ${\cal H}^{(1)}$:
\begin{equation}
{\cal A}=\{A_{\bf m};{\bf m}\in R^3,\vert{\bf m}\vert=1\},
\end{equation}
i.e. we include {\it all}\/ pure state projectors into the set of
observables. A simple choice for completeness relation (1) has the
integral form
$\sum_{\bf m} A_{\bf m} \equiv {1\over2\pi}\int A_{\bf m}d\Omega =1$,
where $d\Omega$ is the solid angle element of ${\bf m}$  \cite{fn1}.
The set of observables is overcomplet. Introduce
Descartes--coordinates on $R^3$, then ${\bf m}=(m_1,m_2,m_3)$.
Let us denote the orthonormal basis vectors
$(1,0,0)$, $(0,1,0)$, $(0,0,1)$
by ${\bf e}_1,{\bf e}_2,{\bf e}_3,$ respectively.
The observables in the POVM $\cal A$ (7) satisfy the constraints
\begin{equation}
A_{\bf m}-\sum_{i=1}^3 m_i A_{{\bf e}_i}
-{1\over2}(1-\sum_{i=1}^3 m_i)I^{(1)}=0
\end{equation}
for all unit vectors ${\bf m}\in R^3$. These constraints could be
rewritten into the general forms (2) if we replaced $I^{(1)}$ by
$\sum_{\bf m^\prime}A_{\bf m^\prime}$
from the normalization condition.

In a similar way we introduce the POVM
\hbox{${\cal B}=\{B_{\bf n};{\bf n}\in R^3,\vert{\bf n}\vert=1\}$} on
${\cal H}^{(2)}$.
Then it follows from our definition of locality proposed earlier in
this Letter that, if the state $\rho$ is local, the LHV model (4) must
exist for the above choices of the POVMs ${\cal A}$ and ${\cal B}$,
i.e.:
\begin{equation}
\langle A_{\bf m}\otimes B_{\bf n} \rangle=\sum_\lambda p_\lambda
E^{(1)}_{\cal A}(A_{\bf m}\vert\lambda)
E^{(2)}_{\cal B}(B_{\bf n}\vert\lambda)
\end{equation}
where
$0\leq E^{(1)}_{\cal A}(A_{\bf m}\vert\lambda)\leq1$,
$0\leq E^{(2)}_{\cal B}(B_{\bf n}\vert\lambda)\leq1$,
and
$\sum_{\bf m}E^{(1)}_{\cal A}(A_{\bf m}\vert\lambda)
=\sum_{\bf n}E^{(2)}_{\cal B}(B_{\bf n}\vert\lambda)=1$,
for all $\lambda$'s.
Let us concentrate on the features of the expectation values
$E^{(1)}_{\cal A}(A_{\bf m}\vert\lambda)$ at fixed hidden
parameter $\lambda$. Given the operator constraints (8),
the consistency condition (6) will take the following form:
\begin{equation}
E^{(1)}_{\cal A}(A_{\bf m}\vert\lambda)
-\sum_{i=1}^3 m_i E^{(1)}_{\cal A}(A_{{\bf e}_i}\vert\lambda)
-{1\over2}(1-\sum_{i=1}^3 m_i)=0.
\end{equation}
It must be satisfied for all unit vectors ${\bf m}\in R^3$.
For notational convenience, we introduce the "polarization vector"
${\bf s}=(s_1,s_2,s_3)\in R^3$ with
$s_i=2E^{(1)}_{\cal A}(A_{{\bf e}_i}\vert\lambda)-1$ for $i=1,2,3$.
One can define the corresponding conditional "density operator":
\begin{equation}
\rho^{(1)}_\lambda\equiv {1\over2}(1+{\bf s}\mbox{\boldmath$\sigma$}).
\end{equation}
The quotation marks are still reminding us that we have not yet proved
the fulfilment of the standard relation $\vert{\bf s}\vert\leq1$
assuring the crucial property $\rho^{(1)}_\lambda\geq0$. If, however,
we substitute $m_1,m_2,m_3$ in Eq.~(10) via the trivial relations
\hbox{$m_i=tr(\sigma_i A_{\bf m})$} the expectation value of
$A_{\bf m}$ can be expressed in the following form:
\begin{equation}
E^{(1)}_{\cal A}(A_{\bf m}\vert\lambda)
=tr( A_{\bf m}\rho^{(1)}_\lambda).
\end{equation}
This equation holds for all pure state projectors $A_{\bf m}$
on ${\cal H}^{(1)}$. Since the LHS is by definition non-negative
the operator $\rho^{(1)}_\lambda$ must be non-negative as well.
Hence one can indeed interpret it as the conditional density operator
of the first subsystem at the given value of the hidden parameter
$\lambda$.

By repeating the same proof for the expectation values
$E^{(2)}_{\cal B}(B_{\bf n}\vert\lambda)$, too, the RHS of
Eq.~(9) can be rewritten as follows:
\begin{equation}
\langle A_{\bf m}\otimes B_{\bf n}\rangle=\sum_\lambda p_\lambda
                                tr(A_{\bf m}\rho^{(1)}_\lambda)
                                tr(B_{\bf n}\rho^{(2)}_\lambda).
\end{equation}
This equation is valid for all pure state projectors
$A_{\bf m},B_{\bf n}$ which leads \cite{fn2} to the following form
for the state $\rho$ of the composite system:
\begin{equation}
\rho=\sum_\lambda
p_\lambda \rho^{(1)}_\lambda)\otimes\rho^{(2)}_\lambda,
\end{equation}
with conditional density operators
$\rho^{(1)}_\lambda,\rho^{(2)}_\lambda$
and with positive normalized probability distribution $p_\lambda$ of
the hidden parameters $\lambda$. This is our central result: local
states are mixtures of product states.

The theorem (14) can also be derived for local states in higher
dimensional Hilbert spaces. We only outline the proof.
If, for instance, $dim{\cal H}^{(1)}=3$ then the "subtle" choice
of POVM ${\cal A}$ means including all one-dimensional Hermitian
projectors $A_\psi=\vert\psi\rangle\langle\psi\vert$ {\it and}\/
all two-dimensional Hermitian projectors
$A_{\psi\phi}
=\vert\psi\rangle\langle\psi\vert+\vert\phi\rangle\langle\phi\vert$
$(\psi,\phi\in{\cal H}^{(1)}; \langle\psi\vert\phi\rangle=0)$
into ${\cal A}$.
There exists the following class of constraints (2):
$A_\psi+A_\phi-A_{\psi\phi}=0$. If a given state $\rho$ is local then,
according to our definitions, a consistent LHV model exists and the
corresponding expectation values are consistent with the above
constraints:
\begin{equation}
 E^{(1)}_{\cal A}(A_\psi\vert\lambda)
+E^{(1)}_{\cal A}(A_\phi\vert\lambda)
-E^{(1)}_{\cal A}(A_{\psi\phi}\vert\lambda)=0.
\end{equation}
This equation holds for all pairs of orthogonal vectors
$\psi,\phi$ of the three-dimensional Hilbert space ${\cal H}^{(1)}$.
Then, Gleason's theorem \cite{Gle,Per} implies the existence of the
density operator $\rho^{(1)}_\lambda$ such that
$E^{(1)}_{\cal A}(A_\psi\vert\lambda)=tr(A_\psi\rho^{(1)}_\lambda)$
and $E^{(1)}_{\cal A}(A_{\psi\phi}\vert\lambda)=
tr(A_{\psi\phi}\rho^{(1)}_\lambda)$.
The proof can obviously be generalized for any finite dimensions of
${\cal H}^{(1)}$ and ${\cal H}^{(2)}$.
Hence the RHS of Eq.~(4) can generally be written as
\begin{equation}
\langle A_\mu \otimes B_\nu \rangle = \sum_\lambda p_\lambda
tr(A_\mu \rho^{(1)}_\lambda) tr(B_\nu \rho^{(2)}_\lambda),
\end{equation}
where $A_\mu$ and $B_\nu$ may be any Hermitian projectors on
${\cal H}^{(1)}$ and ${\cal H}^{(2)}$, respectively. This completes
the proof of Eq.~(14).

We have pointed out that, according to a plausible definition of
local hidden variable models based on generalized measurements,
all local density operators can be decomposed as convex mixtures of
the subsystems' conditional density operators, see Eq.~(14).
If a given state $\rho$ can not be expressed in the form (14) at all
then there exist certain POVMs ${\cal A,B}$ which do {\it not}\/ admit
any consistent LHV model. Logically it means that the statistics of
joint nonideal measurements on $A_\mu$ and $B_\nu$ show nonlocal
correlations and/or violate the linear constraints (6) of consistency.
In fact, however, we have not yet obtained any constructive algorithm
to find these POVMs for a generic non-local (i.e. non-product) state
though the problem has been solved earlier for pure \cite{Gis} as well
as for a special class of mixed non-product states \cite{Pop}. In both
cases the solutions have essentially been based on testing standard
Bell-operators. Further investigations are needed to see what nonideal
measurements can reveal nonlocality of generic non-product mixed
states. The merit of our Letter is the proof of existence of such
sensitive generalized measurements.

\bigskip

This work was supported by the grants OTKA No. 1822/91 and T016047.


\begin{thebibliography}{99}
\bibitem{fn1} Though we use the $SU(2)$-covariant POVM including
              continuum number of observables, forthcoming derivations
              could equally be done with countable or finite number of
              them, due to the compactness of the unitary groups on
              finite-dimensional Hilbert spaces. Proofs, however,
              would then become more circuitous.
\end{thebibliography}
\end{document}